\documentclass[a4paper,fleqn]{cas-dc}
\usepackage{amssymb}
\usepackage{hyperref}
\usepackage{graphicx}
\usepackage[caption=false]{subfig}
\usepackage{lipsum}
\usepackage{lscape}
\usepackage{amsmath}
\usepackage{multirow}
\usepackage[figuresright]{rotating}
\usepackage{lineno}
\usepackage{algorithm} 
\usepackage{algpseudocode} 
\usepackage{comment}
\usepackage{lineno}
\usepackage{nomencl}
\makenomenclature
\usepackage{calc}
\usepackage[T1]{fontenc}
\usepackage{relsize}
\usepackage{subfig}
\usepackage{gensymb}
\usepackage{orcidlink}

\DeclareMathOperator*{\argmax}{arg\,max}

\usepackage[sort,comma,authoryear,round]{natbib}

\setcitestyle{authoryear,open={(},close={)},citesep={;}} 

\hypersetup{
  colorlinks,
  citecolor=Violet,
  linkcolor=Red,
  urlcolor=Blue}

\newcolumntype{P}[1]{>{\centering\arraybackslash}p{#1}}

\def\tsc#1{\csdef{#1}{\textsc{\lowercase{#1}}\xspace}}
\tsc{WGM}
\tsc{QE}
\tsc{EP}
\tsc{PMS}
\tsc{BEC}
\tsc{DE}

\begin{document}
\let\WriteBookmarks\relax
\def\floatpagepagefraction{1}
\def\textpagefraction{.001}
\shorttitle{RL \& LLMs in Crop Management}
\shortauthors{Chen and Huang (2024)}
 
\title [mode = title]{Integrating Reinforcement Learning and Large Language Models for Crop Production Process Management Optimization and Control through A New Knowledge-Based Deep Learning Paradigm}

\author[1]{Dong Chen\orcidlink{0000-0002-0551-5084}}\ead{dc2528@msstate.edu}
\author[2]{Yanbo Huang\orcidlink{0000-0002-1409-8868}*}\ead{Yanbo.Huang@usda.gov}

\address[1]{Department of Agricultural and Biological Engineering, Mississippi State University, MS 39762, USA}
\address[2]{United States Department of Agriculture Agricultural Research Service, Genetics and Sustainable Agriculture Research Unit, MS 39762, USA}

\address{* Yanbo Huang (\textit{Yanbo.Huang@usda.gov}) is the corresponding author}

\begin{abstract}
Efficient and sustainable crop production process management is crucial to meet the growing global demand for food, fuel, and feed while minimizing environmental impacts. Traditional crop management practices, often developed through empirical experience, face significant challenges in adapting to the dynamic nature of modern agriculture, which is influenced by factors such as climate change, soil variability, and market conditions. Recently, reinforcement learning (RL) and large language models (LLMs) bring transformative potential, with RL providing adaptive methodologies to learn optimal strategies and LLMs offering vast, superhuman knowledge across agricultural domains, enabling informed, context-specific decision-making. This paper systematically examines how the integration of RL and LLMs into crop management decision support systems (DSSs) can drive advancements in agricultural practice. We explore recent advancements in RL and LLM algorithms, their application within crop management, and the use of crop management simulators to develop these technologies. The convergence of RL and LLMs with crop management DSSs presents new opportunities to optimize agricultural practices through data-driven, adaptive solutions that can address the uncertainties and complexities of crop production. However, this integration also brings challenges, particularly in real-world deployment. We discuss these challenges and propose potential solutions, including the use of offline RL and enhanced LLM integration, to maximize the effectiveness and sustainability of crop management. Our findings emphasize the need for continued research and innovation to unlock the full potential of these advanced tools in transforming agricultural systems into optimal and controllable ones.
\end{abstract}


\maketitle

\section{Introduction}

\begin{table*}[!ht]
\renewcommand{\arraystretch}{1.4}
\centering
\caption{Nomenclature}
\label{tab:nomen1}
\resizebox{0.98 \textwidth}{!}{%
\begin{tabular}{|ll|ll|}
\hline
\textbf{Nomenclature} &                                                                                                 & LaMDA & Language Model for Dialogue Applications  \citep{ouyang2022training}                                                                  \\ 
RL           & Reinforcement learning                                                                                & LLAMA         &  Large Language Model Meta AI  \citep{touvron2023llama}                                                                                        \\ 
LLMs           & Large language models                                                                                   & LINTUL-3       &  Light INterception and UTilization \citep{shibu2010lintul3}                                                                                              \\ 
DSSs           & Decision support systems                                                                        & DSSAT        & Decision Support System for Agrotechnology Transfer \citep{hoogenboom2019dssat}                                                                                \\ 
PPO          & Proximal Policy Optimization    \citep{schulman2017proximal}                                                               & CGM         & Crop growth model                                                                                                                                          \\ 
VQA          & Visual question answering                                                                       & IL        & imitation learning                                                                       \\ 
DQN          & Deep Q-network  \citep{mnih2015human}                                                                             
     & BCQ         & Batch-constrained deep Q-learning    \citep{fujimoto2019off}                                                                                                                                  \\ 
AI         & Artificial intelligence                                                                            & VAE     & Variational autoencoder \citep{kingma2019introduction} \\ 
MDP        & Markov Decision Process                                                                            & CQL       & Conservative Q-Learning \citep{kumar2020conservative}                                                     \\ 
DDPG         & Deep Deterministic Policy Gradient \citep{lillicrap2015continuous}                                                                     & IQL     & Implicit Q-learning \citep{kostrikov2021offline}                                                                       \\ 
A2C        & Advantage Actor-Critic  \citep{mnih2016asynchronous}                                                             & SAC         & Soft Actor-Critic \citep{haarnoja2018soft}                                                                              \\ 
A3C         & Asynchronous Advantage Actor-Critic \citep{mnih2016asynchronous}                                          & 
     NLP    & Natural language processing                                                                              \\ 
BERT         & Bidirectional Encoder Representations from Transformers \citep{devlin2018bert}                          & PaLM         & Pathways Language Model \citep{wei2022chain}        \\  
GPT-3        & Generative pretrained transformer 3 \citep{brown2020language}                               
      & HER         & Hindsight Experience Replay \citep{andrychowicz2017hindsight}                                                                     \\ 
GPT-4        & Generative pretrained transformer 4 \citep{openai2023gpt4}                                 
     & ChatGPT        & Chat generative pretrained transformer \citep{chatgpt}                                                                                             \\ \hline
\end{tabular}
}
\end{table*}

\label{sec:intro}
The ever-growing world population is intensifying the food security challenge, driving a sharp increase in global demand for agricultural crops used for food, feed, and fuel \citep{ghosh2024exponential, li2023label}. 
However, meeting this demand comes with substantial environmental costs \citep{turchetta2022learning}. For example, while nitrogen (N) is crucial for plant growth, its production as mineral fertilizer and incomplete absorption by plants are major contributors to greenhouse gas emissions \citep{menegat2022greenhouse}, and pollution of surface and groundwater \citep{mateo2018more}. Other practices, such as tillage, can reduce soil organic matter and excessive irrigation can deplete aquifers \citep{havlin1990crop, scanlon2012groundwater}, further highlighting the environmental challenges posed by modern agriculture.

Efficient and sustainable crop production process management  practices can not only enhance productivity and lead to higher yields with improved quality but also mitigate negative environmental impacts \citep{crop_management2021}. Crop management involves a range of agricultural practices aimed at optimizing crop growth, development, and yield. The \textit{timing} and \textit{sequence} of these practices are influenced by various factors, including the type of crop, the intended harvest product, and sowing methods. Additionally, variables such as plant age, soil conditions, and uncertainties like climate and weather patterns play a critical role in determining the best practices. 
However, traditional best practices, developed through empirical experience and academic research, may no longer be as effective in the face of changing climate and market conditions \citep{wu2024new}. This raises concerns regarding the adequacy of current strategies and underscores the need for innovative, efficient, and adaptable crop management decision support systems (DSSs) \citep{skhiri2012impact}. 

Recently, data-driven methods, particularly reinforcement learning (RL), have gained significant attention and have been explored in various applications, including games \citep{silver2016mastering}, robotics \citep{kaufmann2023champion}, and precision agriculture \citep{gandhi2022deep}. RL has also been applied to crop management \citep{gautron2022reinforcement}, with the goal of optimizing agricultural practices while minimizing environmental impact. For instance, in \cite{gautron2022gym}, the Proximal Policy Optimization (PPO) algorithm \citep{schulman2017proximal} was designed to optimize fertilizer and irrigation management strategies and showed that the PPO algorithm ended with the highest mean accumulated return compared to the null and expert policies. In \cite{turchetta2022learning}, the authors applied PPO on a multi-year, multi-crop crop management problem, showing that PPO greatly outperformed the expert recommendations by achieving gains in average yearly profit ranging from 11\% over one-year horizon experiments to almost 30\% over 5 years. Despite these successes, the practical application of RL in crop management faces significant challenges, as all existing RL algorithms for crop management rely on real-time interaction with the environment, which is impractical in real-world scenarios where crop growth spans days to years \citep{levine2020offline}. For instance, perennial crops such as apple trees, grapevines, or coffee plants have growth cycles that span multiple years, requiring long-term management strategies. Even annual crops like wheat or maize involve growth stages that take several months from planting to harvest. The complexity and duration of these growth cycles present significant challenges for applying existing RL algorithms, which typically rely on frequent, real-time interactions with the environment to learn optimal strategies.

On the other hand, large language models (LLMs) have emerged as powerful tools capable of processing and generating natural language, opening new possibilities for integrating expert knowledge and complex data-driven insights into crop management \citep{wu2024new, li2024foundation}. For instance, in \cite{wu2024new}, the authors innovatively combined RL, an LLM, and crop simulations for crop management. Specifically, deep RL, particularly a deep Q-network (DQN) \citep{mnih2015human}, was employed to train management policies by leveraging diverse state variables produced by simulators. These state variables were then converted into more informative language messages, enhancing the LLM's (i.e., DistilBERT \citep{sanh2019distilbert}) ability to understand the states and identify optimal management strategies. The experimental results demonstrated the model's exceptional learning capabilities, achieving state-of-the-art performance across multiple evaluation metrics. Notably, it resulted in an over 49\% improvement in economic profit when simulated on maize crops in Florida (US) and Zaragoza (Spain). However, the LLM was primarily used as a state encoder, focusing on converting raw data into a format suitable for RL algorithms. This approach, while effective, only scratches the surface of the potential applications of LLMs in crop management. 

In this paper, we present a systematic investigation into the integration of RL and LLMs in crop management. We review recent advancements in RL and LLM algorithms and their application within crop management DSSs, as well as the existing crop management simulators used to develop these algorithms. Additionally, we discuss the technical challenges associated with the real-world deployment of RL and LLMs as well as exploring potential solutions, including the application of offline RL and enhanced LLM integration, to enhance the effectiveness and sustainability of these technologies in crop management.

The rest of the paper is organized as follows. Section~\ref{sec2} outlines the problem formulation for crop management, while Section~\ref{sec3} introduces key concepts of RL and LLMs. Section~\ref{sec4} presents an in-depth overview of notable RL algorithms and simulators relevant to crop management. In Section~\ref{sec5}, we discuss the challenges and future directions for integrating RL and LLMs, with a concluding summary provided in Section~\ref{sec6}.

\section{Crop Management DSSs}
\label{sec2}
\begin{figure}[!ht]
  \centering
  \includegraphics[width=0.46\textwidth]{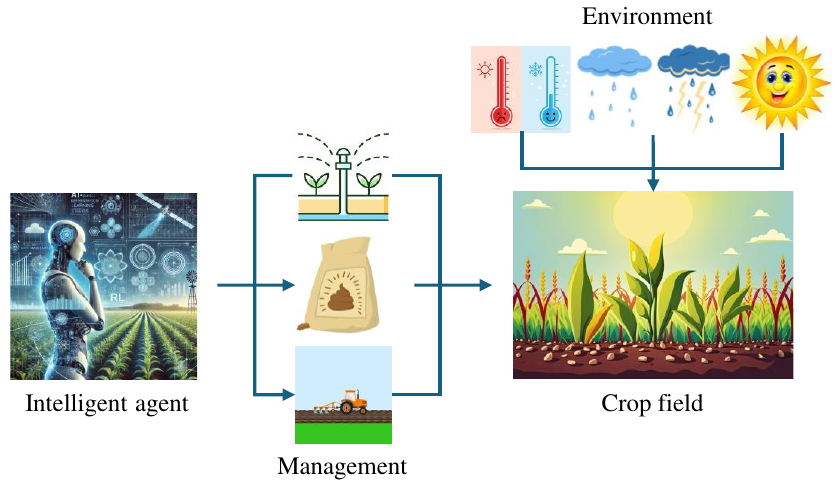}
  \caption{Overview of a crop management system, highlighting the interaction between an intelligent crop management agent, environmental factors (e.g., weather conditions), and the crop field. The agent analyzes environmental conditions and makes management decisions to optimize crop growth and yield.}
  \label{fig:crop_DSS}
\end{figure}

Mastering sustainable crop management practices is a complex task. As shown in Figure~\ref{fig:crop_DSS}, agricultural processes involve the dynamic interaction of complex physical, chemical, and biological factors \citep{husson2021soil}. The evolution of these systems is further complicated by uncertain variables such as weather events, soil conditions, pest infestations, and human interventions, all contributing to the variability and complexity of crop growth and development \citep{gautron2022gym}. For instance, irrigation scheduling must account for future weather conditions to ensure that plants receive adequate water without overuse, which could result in soil erosion or nutrient leaching. Similarly, effective pest management requires precise timing and treatment selection to minimize crop damage while preventing pesticide resistance \citep{darshna2015smart, dent2020insect}. Therefore, the \textit{timing} and \textit{sequence} of these practices are critical to ensuring optimal crop health and sustainable agricultural outcomes.

Crop management DSSs are designed to optimize agricultural practices by providing data-driven recommendations that enhance both crop productivity and sustainability, while accounting for the inherent complexities of agricultural systems \citep{turchetta2022learning}. These systems commonly support various aspects of crop management, including fertilization, irrigation, pest and disease control, weed management, and more. The end users of these DSSs can range from researchers and local advisers to farmers, depending on the system's complexity and objectives \citep{gautron2022reinforcement}. The versatility of crop management DSSs lies in their ability to integrate vast amounts of data from various sources, such as weather forecasts, soil sensors, crop models, and historical yield data. By analyzing these data inputs, DSSs can generate actionable insights and recommendations tailored to the specific conditions of each field \citep{gautron2022reinforcement}. For instance, advanced DSSs can predict the optimal timing for fertilizer application by considering current soil nutrient levels, weather predictions, and the growth stage of the crop. 

The crop management problem can be mathematically formulated as an optimization problem where the objective is to maximize crop yield $\mathcal{Y}$, while minimizing the use of resources $\mathcal{R}$ (such as water, fertilizer, and pesticides) and environmental impact $\mathcal{E}$. This can be expressed as:
\begin{subequations}\label{crop_dss}
\begin{align}
\max_{X_t,U_t} \quad & \mathcal{Y} =  f(s,a,m)\\
\min_{X_t,U_t} \quad & \mathcal{R} =  g(w,f,p)\\
\min_{X_t,U_t} \quad & \mathcal{E} =  h(n,l,g)\\
\textrm{s.t.} \quad & R_i \leq R_{i, \text{max}}, \text{ for each resource } i\\
  & E_j \leq E_{j, \text{max}}, \text{ for each environmental factor } j\\
  & s_{t+1} \leftarrow \Phi(s_t, a_t, m_t) + \xi_t
\end{align}
\end{subequations}
where $s$ represents the state variables, including soil conditions, weather conditions, and plant health. $a$ represents the set of control variables or actions taken, such as irrigation levels, fertilizer application rates, and pesticide use. $m$ represents management practices, including crop rotation schedules and tillage practices.
$w$, $f$, and $p$ are the resources used, namely water, fertilizer, and pesticides, respectively. $n$, $l$, and $g$ represent the environmental factors to be minimized, such as nutrient runoff, loss of biodiversity, and greenhouse gas emissions. The functions $f(s, a, m)$, $g(w, f, p)$, and $h(n, l, g)$ capture the relationships between the state of the crop management system, the actions taken, and the resulting yield, resource usage, and environmental impact. 

Additionally, the system is subject to various constraints Eq.(1d-f). $\Phi$ represents the dynamic evolution of the state variables based on the actions taken and management practices, and $\xi_t$ represents stochastic disturbances (e.g., unexpected weather events). The goal of a crop management DSS is to provide the optimal set of actions $a^*$ and management practices $m^*$ that maximize yield while adhering to resource and environmental constraints:
\begin{equation}
    (a^*, m^*) = \arg\max_{a, m} [ f(s, a, m)] \text{, subject to constraints.}
\end{equation}
By solving this optimization problem, a DSS can generate recommendations that guide farmers in making informed decisions that balance productivity with sustainability.

\section{Advancements in RL and LLMs}
\label{sec3}
The intersection of Artificial Intelligence (AI) and cybernetics continues to forge transformative pathways across various agricultural domains, reshaping our approach to problem-solving and system optimization \citep{huang2021agricultural}. This section delves into the recent breakthroughs and fundamental technologies driving this evolution, with a special emphasis on RL and LLMs.

\subsection{Preliminaries of RL}
RL is often formulated within the framework of a Markov Decision Process (MDP), making it a powerful approach for data-driven and sequential decision-making \citep{kaelbling1996reinforcement}.
Recently, deep neural networks (DNN) have significantly enhanced the capacity of RL to address complex challenges \citep{mnih2015human}. Key developments in this field include algorithms such as the Deep Deterministic Policy Gradient (DDPG) \citep{lillicrap2015continuous},  Deep Q-Network (DQN) \citep{mnih2015human}, and Advantage Actor-Critic (A2C) \citep{mnih2016asynchronous}. For example, AlphaStar, which employs techniques similar to DQN, achieved a notable breakthrough by outperforming professional players in the strategic game StarCraft II \citep{vinyals2019grandmaster}, highlighting RL's capabilities in decision-making under uncertainty. Moreover, A2C has been instrumental in developing autonomous driving systems, where it enables vehicles to navigate environments with dynamic obstacles effectively \citep{zhou2022multi}.

\begin{figure}[!ht]
  \centering
  \includegraphics[width=0.45\textwidth]{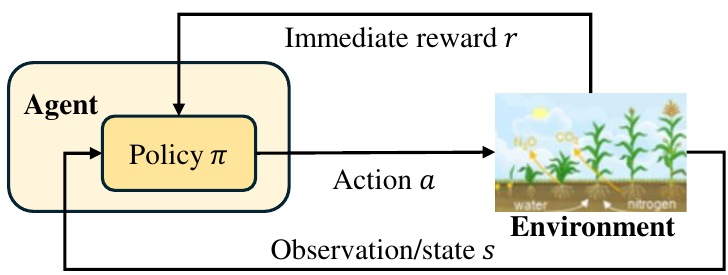}
  \vspace{1pt}
  \caption{Framework of RL for crop management.
  }
  \label{fig:rl}
\end{figure}

Within an RL framework (depicted in Fig.~\ref{fig:rl}), an \textbf{agent} (e.g., a crop management DSS) learns to operate within an \textbf{environment} (e.g., a field or greenhouse) through trial and error. The agent observes the \textbf{state} $s$  (e.g., soil moisture levels and plant growth stages) of the environment, performs specific \textbf{actions} $a$ (e.g., deciding whether to irrigate, fertilize, or apply pesticides), and receives feedback in the form of \textbf{rewards} $r$ (e.g., increased crop yield or reduced water usage) and new state. This feedback mechanism allows the agent to evaluate the effectiveness of its actions and adjust its strategy accordingly. The objective of the RL agent is to learn an optimal \textbf{policy} $\pi^{*}: \mathcal{S} \rightarrow \mathcal{A}$, which maps states to actions in a way that maximizes the accumulated reward, defined as $R_t = \sum_{k=0}^{T} \gamma ^k r_{t+k}$. Here, $r_{t+k}$ represents the reward received at time step $t+k$, and $\gamma \in (0,1]$ is the discount factor, which determines the weight assigned to future rewards relative to immediate ones \citep{chen2023deep}.

Many RL problems are modeled as MDPs, providing a structured approach to decision-making where outcomes are influenced by both the environment and the agent's actions. An MDP is formally represented as $\mathcal{M} = (\mathcal{S}, \mathcal{A}, \mathcal{R}, \mathcal{P})$, comprising states, actions, rewards, and transition dynamics:
\begin{enumerate}
\item \textbf{State space} $\mathcal{S}$: Represents the environment's condition at any given time $t$, including all necessary details for the agent's decision-making. An \textbf{observation} $o$ may provide a partial view of the state; in fully observed environments, the observation matches the state, whereas in partially observed environments, it might contain incomplete information.

\item \textbf{Action space} $\mathcal{A}$: Defines the set of possible actions or decisions the agent can make based on its current state. The set of actions can vary by state and may be either discrete or continuous.

\item \textbf{Reward function} $\mathcal{R}(s_t, a_t, s_{t+1})$: A scalar feedback signal provided by the environment after the agent takes action $a_t$ in state $s_t$. The agent's goal is to maximize the total accumulated reward over time.

\item \textbf{Transition probability} $\mathcal{P}_{ss'}(S_{t+1}=s' | S_{t}=s)$: The probability of transitioning from one state to another as a result of the agent's actions.
\end{enumerate}

The following subsections will dive deeper into three major categories of RL algorithms: value-based methods, policy-based methods, and actor-critic methods.

\subsubsection{Value-Based Methods}
In value-based RL, a key concept is the Q-function, represented as $Q_{\theta}$, which estimates the expected cumulative reward of taking a particular action $a$ in a given state $s$ and subsequently following the optimal policy. It is parameterized by a set of variables $\theta$, allowing for various function approximators, such as conventional Q-tables \citep{watkins1992q} and more complex models like deep neural networks (DNNs) \citep{mnih2015human}.
The primary mechanism for updating these parameters is based on the temporal difference (TD) error, expressed as $(\mathcal{T}Q_{\theta^-} - Q_{\theta})(s_t, a_t)$, where $\mathcal{T}$ denotes the dynamic programming operator. Additionally, $Q_{\theta^-}$ refers to a ``frozen'' copy of the model parameters $\theta^-$, which remains fixed during updates to enhance stability \citep{chu2019model}. To reduce the variance of Q-value estimates and improve the algorithm's exploration efficiency, techniques like the $\epsilon$-greedy strategy and experience replay are commonly incorporated within deep Q-learning approaches \citep{szepesvari2022algorithms}. The optimal action 
$a^*(s)$ is determined by: 
\begin{equation}\label{eqn:optimal_a}
a^*(s) = \argmax_a{Q^*(s_t=s, a_t=a)},
\end{equation}

Some widely implemented deep Q-learning-based algorithms include DQN \citep{mnih2013playing}, Hindsight Experience Replay (HER) \citep{andrychowicz2017hindsight}, and DDQN \citep{van2016deep}.

\subsubsection{\textbf{Policy-Based Methods}}
In contrast to value-based approaches like Q-learning, policy-based methods directly optimize the policy, $\pi_\theta$, through its own set of parameters $\theta$.
The primary objective of adjusting these parameters is to increase the likelihood of actions that lead to higher accumulated rewards. This optimization is carried out through a predefined loss function:
\begin{equation}\label{eqn:policy_loss}
\nabla_{\theta} \mathcal{L} (\pi_{\theta}) =\mathop{\mathbb{E}}_{\tau \sim \pi_{\theta}}[\sum_{t=0}^T \nabla_{\theta} \log \pi_{\theta} (a_t|s_t) R_t],
\end{equation}
where the parameters of the policy network are refined using stochastic gradient ascent:
\begin{equation}\label{eqn:gradient_ascent}
\theta_{k+1} = \theta_{k} + \alpha \nabla_{\theta} \mathcal{L} (\pi_{\theta}),
\end{equation}

Compared to Q-learning, policy gradient approaches exhibit a stronger ability to handle non-stationary dynamics within individual trajectories but can experience higher outcome variance \citep{chu2019multi}. Notable algorithms built on policy gradients include Deep Deterministic Policy Gradient (DDPG) \citep{lillicrap2015continuous} and PPO \citep{schulman2017proximal}.

\subsubsection{Actor-Critic Methods}
Actor-critic approaches aim to mitigate the high variance commonly associated with policy gradient methods by incorporating an advantage function, which enhances policy updates by integrating both policy optimization (the actor) and value estimation (the critic) \citep{mnih2016asynchronous}. The advantage function, which plays a central role in this approach, is defined as:
\begin{equation}\label{eqn:advantage_function}
A^\pi (s_t, a_t) = Q^{\pi_\theta}(s_t, a_t) - V_w(s_t),
\end{equation}
where the update of parameters $\theta$ is guided by a policy loss function articulated as:
\begin{equation}\label{eqn:advantage_function_loss}
\nabla_{\theta} \mathcal{L}=\mathop{\mathbb{E}}_{\pi_{\theta}}[\sum_{t=0}^T \nabla_{\theta} \log \pi_{\theta} (a_t|s_t) A_t],
\end{equation}

Simultaneously, the critic updates the value function with the following objective:
\begin{equation}\label{eqn:advantage_function_value_loss}
\mathcal{L}=\min_{w} \mathop{\mathbb{E}}_{\mathcal{D}}[(R_t + \gamma V_{w^-}(s_t) - V_w(s_t))^2 ],
\end{equation}
where $\mathcal{D}$ refers to the experience replay buffer that stores past experiences. This buffer, coupled with a target network using parameters from prior iterations, facilitates stable learning \citep{chen2023deep}. Notable actor-critic algorithms include Soft Actor-Critic (SAC) \citep{haarnoja2018soft} and Asynchronous Advantage Actor-Critic (A3C) \citep{mnih2016asynchronous}.

\subsection{Preliminaries of LLMs}
In recent years, the field of computational natural language processing (NLP) has undergone a transformation with the advent of LLMs \citep{devlin2018bert, brown2020language}. Leveraging the power of deep learning, LLMs have revolutionized NLP tasks by modeling and understanding language on an unprecedented scale. These models are typically built on DNNs that can learn context and meaning from large amounts of text data, enabling them to perform tasks ranging from language translation and text summarization to question-answering and dialogue generation, and beyond \citep{li2023foundation}. 

One of the key milestones in the rise of LLMs has been the development of Transformer-based architectures \citep{vaswani2017attention}, which have paved the way for more powerful and versatile language models.
A prominent example of this innovation is the \underline{B}idirectional \underline{E}ncoder \underline{R}epresentations from \underline{T}ransformers (BERT, \citep{devlin2018bert}), which comes in two variants: $\text{BERT}_\text{base}$ with 110M parameters and $\text{BERT}_\text{large}$ with 340M parameters. By simply adding an output layer for task-specific fine-tuning, BERT can be adapted for a variety of NLP tasks like question answering and natural language inference. This adaptability allows BERT to achieve superior performance across tasks without major alterations to the model's architecture, often surpassing task-specific designs.
Following the success of BERT, Google introduced the \underline{P}athways \underline{L}anguage \underline{M}odel (PaLM) \citep{wei2022chain}, followed by the Language Model for Dialogue Applications (LaMDA) \citep{thoppilan2022lamda}, PaLM2 \citep{anil2023palm}, and, most recently, the advanced LLMs, Gemini and Gemma 2 (formerly known as Google Bard) \citep{team2023gemini, team2024gemma}. PaLM, with 540 billion parameters, and its successor PaLM2, featuring 340 billion parameters, are built as dense decoder-only Transformer models and trained using the Pathways system \citep{barham2022pathways} for efficient learning. Gemma 2 offers multiple model sizes, including 2B, 9B, and 27B parameters, further expanding the capabilities of Google's LLMs. 

OpenAI has also made significant strides in developing LLMs, introducing the \underline{G}enerative \underline{P}retrained \underline{T}ransformer series: GPT-2 \citep{radford2019language}, GPT-3 \citep{brown2020language}, GPT-4 \citep{openai2023gpt4}, GPT-4o\footnote{GPT-4o: \url{https://openai.com/index/hello-gpt-4o/}}, and OpenAI o1\footnote{OpenAI o1: \url{https://openai.com/o1/}}. 
Each iteration has substantially increased in scale, from 1.5 billion parameters in GPT-2 to 175 billion in GPT-3. GPT-4o boasts over 200 billion parameters, surpassing the 175 billion parameters of GPT-4, setting new benchmarks in language understanding and generation. 
A specialized version, ChatGPT \citep{chatgpt}, has been fine-tuned to facilitate human-like conversational interactions. By modifying the output layer, the pre-trained GPT models can be tailored for various tasks, such as text generation, translation, and dialogue systems, frequently outperforming more narrowly focused models while maintaining a unified foundational architecture.

More recently, Claude 3 and Claude 3.5, developed by Anthropic\footnote{Claude ai website: \url{https://www.anthropic.com/claude}.}, represents a state-of-the-art advancement in LLM systems. It showcases capabilities across various complex tasks, encompassing domains such as undergraduate-level knowledge (MMLU), graduate-level reasoning (GPQA), and basic mathematics (GSM8K). Claude 3 displays near-human comprehension and fluency, positioning it at the forefront of general intelligence research \citep{Anthropic_Claude3}. 
Meta (formerly Facebook AI) has also been a significant contributor to the development of LLMs. Meta introduced the \underline{L}arge \underline{L}anguage \underline{M}odel \underline{M}eta \underline{A}I (LLAMA) \citep{touvron2023llama}, designed to enhance efficiency and performance across various language tasks. LLAMA was notable for its parameter efficiency, demonstrating high performance with fewer parameters compared to other LLMs of similar capabilities. Building on this foundation, Meta released LLAMA 2 \citep{touvron2023llama} and LLAMA 3 \citep{dubey2024llama}, which further refined the model’s capabilities, making it more effective in real-world applications such as text generation, question-answering, and dialogue systems.

\begin{figure*}[]
  \centering
  \includegraphics[width=0.8\textwidth]{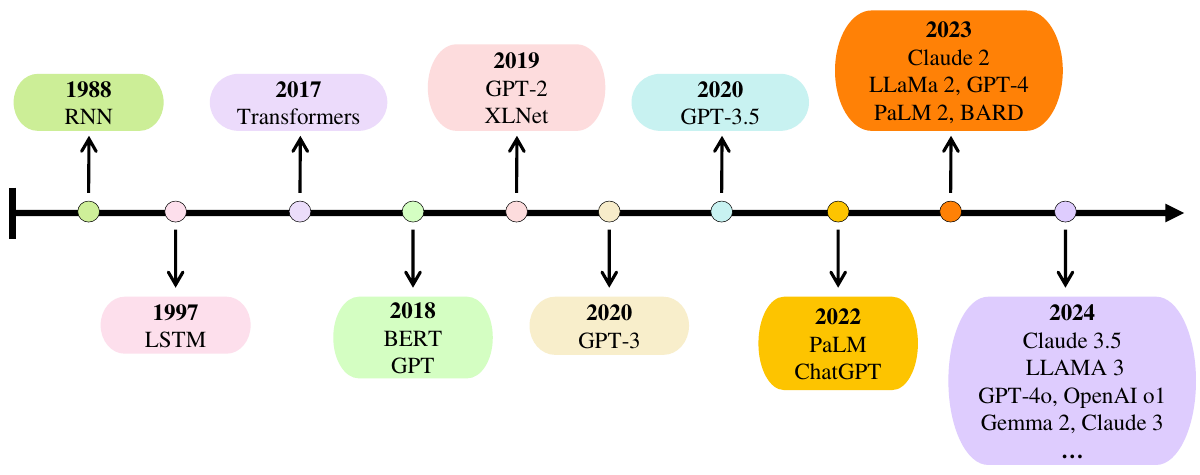}
  \caption{Millionstone of LLMs.
  }
  \label{fig:LLM_timeline}
\end{figure*}

Figure~\ref{fig:LLM_timeline} shows the timeline of the LLMs. Together, these advances in LLM development highlight a significant leap toward creating highly efficient, versatile, and accessible language models. Their collective progress not only shows cutting-edge performance on diverse benchmarks but also emphasizes interpretability, usability, and alignment with human values, establishing these models as foundational tools for AI research and a wide range of practical applications, such as precise crop management.

\section{Current Progress in Crop Management DSSs}
\label{sec4}
In this section, we review the state-of-the-art developments in crop management DSSs, highlighting important research and the simulators that play a critical role in advancing this domain. By exploring both theoretical developments and practical tools, we gain a comprehensive understanding of how modern technologies are being integrated into agricultural practices to enhance productivity, sustainability, and precision in crop management. 

\subsection{Recent Advances in Crop Management DSSs}
Recent advancements in crop management DSSs have demonstrated significant potential in optimizing agricultural practices through RL. A notable example is presented in \cite{overweg2021cropgym}, where the authors employed the PPO algorithm to optimize fertilizer management strategies using a process-based crop growth model, namely Light INterception and UTilization (LINTUL-3) \citep{shibu2010lintul3}. The MDP is formulated as $\mathcal{M} = (\mathcal{S}, \mathcal{A}, \mathcal{P}, \mathcal{R})$ with components defined as follows: the state space $\mathcal{S}$ consists of the crop's current state ($s_{\text{crop}}$) and multidimensional weather observations ($s_{\text{weather}}$). The action space $\mathcal{A}$ is discretized as different fertilizer doses, while the reward function $\mathcal{R}$ balances yield gains and environmental impact by comparing mass gains with and without fertilizer application and penalizing for fertilizer usage. The state transitions ($\mathcal{P}$) are determined by the deterministic crop model and weather sequences. The evaluation of the algorithm across multiple growing seasons (1984, 1994, 2004, and 2014) demonstrated that the PPO agent tended to improve fertilization, resulting in promising yields in some years compared to other agents.

The work of \cite{gautron2022gym} introduced an RL environment, gym-DSSAT, which is built upon the widely used crop model Decision Support System for Agrotechnology Transfer (DSSAT) \citep{hoogenboom2019dssat}. This simulator addresses three tasks: fertilization-only, irrigation-only, and a mixed fertilization and irrigation task. For the fertilization-only task, the MDP $\mathcal{M}_f$ has a state space $\mathcal{S}$ that captures crop stages and daily rainfall data, and an action space that specifies daily nitrogen fertilization. The irrigation-only problem’s MDP $\mathcal{M}_i$ focuses on soil moisture and water management states, with an action space defining daily irrigation amounts. The authors used the PPO algorithm to learn optimal strategies for maize fertilization and irrigation, achieving superior accumulated returns compared to null and expert policies.

Similarly, \cite{turchetta2022learning} developed an RL environment based on Cycles \citep{kemanian2022cycles}, a multi-year, multi-crop crop growth model (CGM) simulator. The MDP $\mathcal{M}$ here incorporates soil, crop, and weather variables as part of the state space. The action space covers a wide range of crop management decisions, such as fertilization, irrigation, planting, and tillage. The simulator's reward function is designed around profitability, driving the adoption of beneficial management practices. In experiments, PPO significantly outperformed expert recommendations, with profit gains ranging from 11\% in one-year experiments to nearly 30\% over five years.

In \cite{tao2022optimizing}, a system was introduced to optimize maize management by simultaneously focusing on nitrogen fertilization and irrigation using RL and imitation learning (IL). The MDP $\mathcal{M}$ comprises soil, crop, and weather variables as state components. The action space includes the amount of nitrogen fertilizer and irrigation water. DQN was utilized to train RL agents in a fully observed setting, with the resulting policies serving as expert guidance for partial observation scenarios via IL. Validated on simulated environments for Florida (1982) and Zaragoza, Spain (1995), the RL-trained policies achieved significant profit gains of 45\% and 55\%, respectively, compared to baseline policies.

An open-source farming environment, Farm-gym, was introduced in \cite{maillard2023farm} for sequential decision-making using RL. The MDP $\mathcal{M}$ integrates multiple components of the agroecosystem, including weather, soil, crops, pollinators, and weeds. Farm-gym provides 17 state variables related to crop and environmental conditions, and 9 actions ranging from weeding to watering. The reward function encompasses factors like biodiversity, resource usage, soil health, yield, and plant development.

In a recent study, \cite{wu2024new} proposed an innovative crop management framework integrating LLMs and RL. The approach uses DQN to train crop management policies based on simulator-derived observations, which are subsequently transformed into more informative language for enhanced interpretability by LLMs. The MDP is built upon the paradigm set in earlier works \citep{tao2022optimizing, wu2022optimizing}, demonstrating the effectiveness of LLM-RL integration in optimizing crop management.

Table~\ref{tab:papers} summarizes the recent efforts in applying RL and LLMs to crop management DSSs, detailing the crop focus, MDP components, and simulators used.

\begin{table*}[!ht]
\centering
\renewcommand{\arraystretch}{1.3}
\caption{Summary of recent research applying RL and LLMs in crop management.} \label{tab:papers}
\resizebox{0.97\textwidth}{!}{
\begin{tabular}{|c|c|ccc|c|}
\hline
\multirow{2}{*}{Paper} & \multirow{2}{*}{Crop} & \multicolumn{3}{c|}{MDP Formulation} & \multirow{2}{*}{Simulator} \\ 
 &  & \multicolumn{1}{c}{Action} & \multicolumn{1}{c}{State} & Reward &  \\ \hline \hline
\cite{overweg2021cropgym} & winter wheat & \multicolumn{1}{c|}{fertilizer} & \multicolumn{1}{c|}{crop data, weather} & yield, fertilizer use & CropGym \\ 
\cite{gautron2022gym} & maize & \multicolumn{1}{c|}{fertilizer, irrigation} & \multicolumn{1}{c|}{crop data, rain} & yield, fertilizer use & Gym-DSSAT \\ 
\cite{turchetta2022learning} & maize, soy & \multicolumn{1}{c|}{fertilizer, irrigation} & \multicolumn{1}{c|}{crop data, weather} & yield, fertilizer use & CYCLESGYM \\ 
\cite{tao2022optimizing} & maize & \multicolumn{1}{c|}{fertilizer, irrigation} & \multicolumn{1}{c|}{crop data, rain} & yield, fertilizer use & Gym-DSSAT \\ 
\cite{wu2022optimizing} & maize & \multicolumn{1}{c|}{fertilizer, irrigation} & \multicolumn{1}{c|}{crop data, rain} & yield, fertilizer use & Gym-DSSAT \\ 
\cite{maillard2023farm} & maize, soy, tomato & \multicolumn{1}{c|}{planting, weeding, watering, fertilizing, etc} & \multicolumn{1}{c|}{crop data, weather} & yield, fertilizer use & Farm-gym \\ 
\cite{wu2024new} & maize & \multicolumn{1}{c|}{fertilizer, irrigation} & \multicolumn{1}{c|}{crop data, rain} & yield, fertilizer use & Gym-DSSAT \\ \hline
\end{tabular}}
\end{table*}

\subsection{Simulators for Crop Management DSSs}
\begin{figure*}[!ht]
  \centering
  \includegraphics[width=0.7\textwidth]{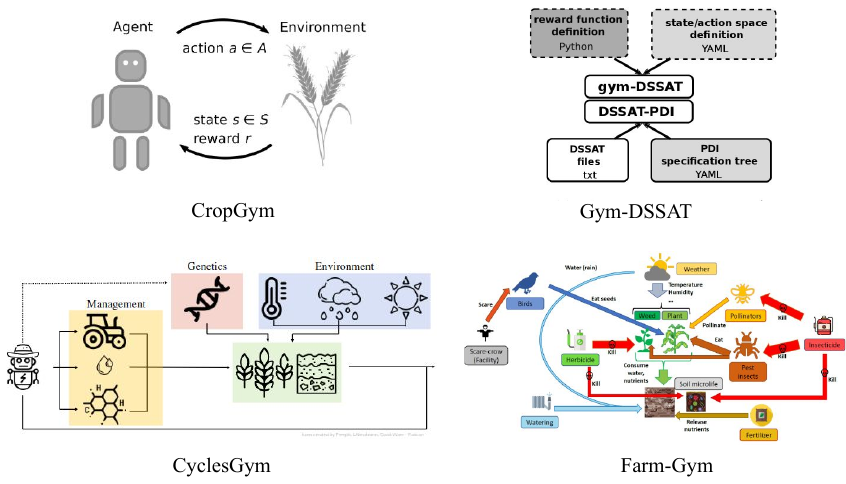}
  \vspace{1pt}
  \caption{Simulators for crop management DSSs, including CropGym \citep{overweg2021cropgym}, CyclesGym \citep{turchetta2022learning}, Gym-DSSAT \citep{hoogenboom2019dssat}, and Farm-Gym \citep{maillard2023farm}.
  }
  \label{fig:simulators}
\end{figure*}

The recent advancements in crop management DSSs have been significantly supported by the development and enhancement of various simulators. These tools play a crucial role in testing and optimizing agricultural decision-support strategies. In this section, we discuss four leading simulators: CropGym, CyclesGym, Gym-DSSAT, and Farm-Gym, which are illustrated in Figure~\ref{fig:simulators}.

\subsubsection{CropGym}
CropGym \citep{overweg2021cropgym} is an advanced simulation environment developed at Wageningen University, designed to optimize fertilizer management in agriculture using RL. Built within the OpenAI Gym framework, it uses process-based crop growth models like LINTUL-3 \citep{shibu2010lintul3} to train RL agents, specifically with the PPO algorithm \citep{schulman2017proximal}. The core aim of CropGym is to identify fertilization policies that maximize crop yield while minimizing environmental impacts. Key designs of the simulator include:
\begin{enumerate}
    \item \textit{State space}: The state space includes outputs from the crop growth model and weather data, providing a realistic and dynamic simulation of agricultural conditions.
    \item \textit{Action space}: The agent can choose from a range of discrete fertilizer doses, exploring various fertilization strategies.
    \item \textit{Reward function}: The reward structure is designed to maximize yield and penalize overuse of fertilizers, thereby promoting environmentally sustainable practices.
\end{enumerate}

CropGym effectively balances agricultural productivity with sustainability, demonstrated by outperforming traditional fertilization approaches in tests. It provides a practical tool for farmers to optimize fertilizer use, reducing environmental footprints and improving profitability. CropGym is utilized in \cite{overweg2021cropgym}.

\subsubsection{CyclesGym}
CyclesGym builds on the Cycles crop growth model and offers an RL environment geared toward long-term and multi-year strategic planning in agriculture \citep{turchetta2022learning}. It provides diverse state spaces and complex action sets covering various aspects of agricultural management, including fertilization, irrigation, and tillage. Its modular design allows users to customize and extend state spaces and actions for different scenarios. Key features include:
\begin{enumerate}
    \item \textit{State space}: Comprising comprehensive environmental, soil, and weather data, it reflects the multifaceted nature of agricultural systems.
    \item \textit{Action space}: Various agricultural actions are included, each with multiple parameters, ranging from planting to fertilization, irrigation, and tillage.
    \item Transitions: CyclesGym simulates state transitions by accounting for weather variations and biological processes in crops and soil, providing realistic scenario modeling.
\end{enumerate}

The modularity of CyclesGym makes it valuable for researchers testing new RL algorithms and agronomists optimizing real-world farming practices. It is compatible with the OpenAI Gym interface, making it accessible to a broad RL community and supported by robust documentation and tutorials. CyclesGym is utilized in \cite{turchetta2022learning}.

\subsubsection{Gym-DSSAT}
Gym-DSSAT is an RL environment based on the well-established Decision Support System for Agrotechnology Transfer (DSSAT) \citep{hoogenboom2019dssat}, a comprehensive crop model developed over three decades. Integrated into the OpenAI Gym interface, Gym-DSSAT provides a sophisticated platform for simulating and optimizing crop management strategies, particularly for maize. Its features include:
\begin{enumerate}
    \item \textit{State space} and \textit{action space}: Gym-DSSAT represents the crop environment in detail, including variables for crop stages, soil conditions, and weather. Actions involve daily decisions on fertilization and irrigation, enabling fine-grained control over crop management.
    \item \textit{MDP formulation}: The environment is framed as a MDP with well-defined state spaces, action spaces, and rewards, allowing precise modeling of agricultural decisions and their impacts on crop outcomes.
    \item \textit{Weather and crop modeling}: By incorporating stochastic weather simulations and detailed crop growth models, Gym-DSSAT enables realistic planning and decision-making under uncertain conditions.
\end{enumerate}

Gym-DSSAT's flexibility allows users to simulate various management strategies and conditions, making it a powerful tool for hypothesis testing and training RL algorithms on diverse datasets. It supports long-term strategy optimization and helps reduce environmental impacts while improving crop yields. Gym-DSSAT is widely utilized in \cite{gautron2022gym, tao2022optimizing, wu2022optimizing, wu2024new}.

\subsubsection{Farm-Gym}
Farm-Gym is an open-source, modular RL platform for simulating and optimizing farm management strategies \citep{maillard2023farm}. It models a farm as a dynamic system with multiple interacting entities and adopts a gamified approach to simulate complex agricultural environments. Key features include:
\begin{enumerate}
    \item \textit{Modular design}: Farm-Gym's architecture enables users to create and customize farm scenarios, studying specific agricultural dynamics, including interactions between entities like plants, pests, weeds, and pollinators.
    \item \textit{Stochastic games}: The platform incorporates real-world unpredictability by using variable weather data and growth models to simulate agricultural challenges.
    \item \textit{Customizable environments}: Users can adjust settings via YAML files, modifying actions, rewards, and conditions to suit specific research or educational objectives.
\end{enumerate}

Farm-Gym also includes an interaction model that captures the dynamics among different farm entities, providing a deeper understanding of complex agricultural ecosystems. By employing RL algorithms, Farm-Gym supports the development of advanced decision-making strategies to optimize inputs like water, fertilizers, and pesticides while considering environmental impacts. It serves as a valuable resource for both researchers and educators seeking to apply machine learning to real-world agricultural challenges. Farm-Gym is utilized in \cite{maillard2023farm}.

Recent advancements in RL for crop management DSSs have shown significant promise in optimizing agricultural practices. By developing advanced simulators and innovative RL algorithms, researchers have effectively tackled key challenges in crop growth modeling, including fertilization, irrigation, and yield optimization. Moreover, the integration of LLMs has further improved the interpretability and overall effectiveness of these systems. However, while RL has shown promise in crop management applications, it also faces several challenges, which will be explored in the next section.

\section{Offline RL \& LLMs for Crop Management}
\label{sec5}
Despite recent advancements in RL and promising results reported in existing studies, most approaches are limited to simulation environments, presenting great challenges for their application to real-world crop management \citep{gautron2022reinforcement}. One of the key issues is \textit{sample inefficiency}—the need for a large number of online interactions with the environment to learn an optimal policy, which is often impractical in agricultural settings due to the time, cost, and seasonal constraints involved in data collection \citep{yu2018towards}. Another challenge is the \textit{design of the reward function}, which must accurately capture the complex objectives of crop management, such as balancing yield maximization with resource conservation \citep{icarte2022reward}. Additionally, the learned policies need to exhibit strong \textit{generalization capabilities} to handle diverse and unforeseen scenarios in real-world agriculture, as variability in weather, soil conditions, and pest dynamics can significantly affect outcomes \citep{li2017deep}.

Offline RL and LLMs offer promising solutions to these challenges. Offline RL enables policy learning from pre-collected datasets, bypassing the need for costly and time-consuming real-world interactions, while LLMs can assist in interpreting complex agricultural data and generating expert knowledge to support decision-making \citep{cao2024survey}. The integration of offline RL and LLMs has the potential to create a synergistic effect, where RL-driven strategies are enhanced by the contextual understanding and adaptability of LLMs. Although research on the application of these techniques to crop management is still emerging, the following subsections will delve into their potential, exploring relevant studies and their implications for developing robust, data-driven crop management strategies.
 
\subsection{Offline RL for Crop Management}
The current RL approaches mainly rely on the online learning paradigm, presenting a significant barrier to adoption in crop management. These methods iteratively gather experience by interacting with the environment using the latest learned policy and subsequently use this experience to improve the policy \citep{ernst2024introduction}. However, in many agricultural contexts, particularly crop management, such online interaction is impractical due to the high costs and time demands of data collection. Even in plant species where online interaction is feasible, it is often preferable to leverage previously collected data instead \citep{levine2020offline}. Offline RL effectively addresses these issues by enabling learning from pre-existing datasets without requiring additional costly and time-consuming interactions with the environment \citep{turchetta2022learning}.

The offline RL problem is a data-driven variant of the traditional RL problem, aiming to optimize the expected return. In the offline RL framework, the agent cannot interact with the environment to collect additional transitions using the behavior policy. Instead, it learns the best policy from a static dataset of transitions, $\mathcal{D} = { (s^i_t, a^i_t, s^i_{t+1}, r^i_t) }$. This approach closely resembles supervised learning, treating $\mathcal{D}$ as the training set for the policy \citep{levine2020offline}. Essentially, the offline RL algorithm must develop a comprehensive understanding of the underlying dynamics of the MDP from this fixed dataset, and then construct a policy $\pi(a|s)$ that maximizes the cumulative reward when deployed. In the context of crop management, offline RL can be particularly useful for developing irrigation strategies, pest control policies, or fertilization schedules without the need for costly and time-consuming real-world experimentation.

\begin{figure*}[!ht]
  \centering
  \includegraphics[width=0.8\textwidth]{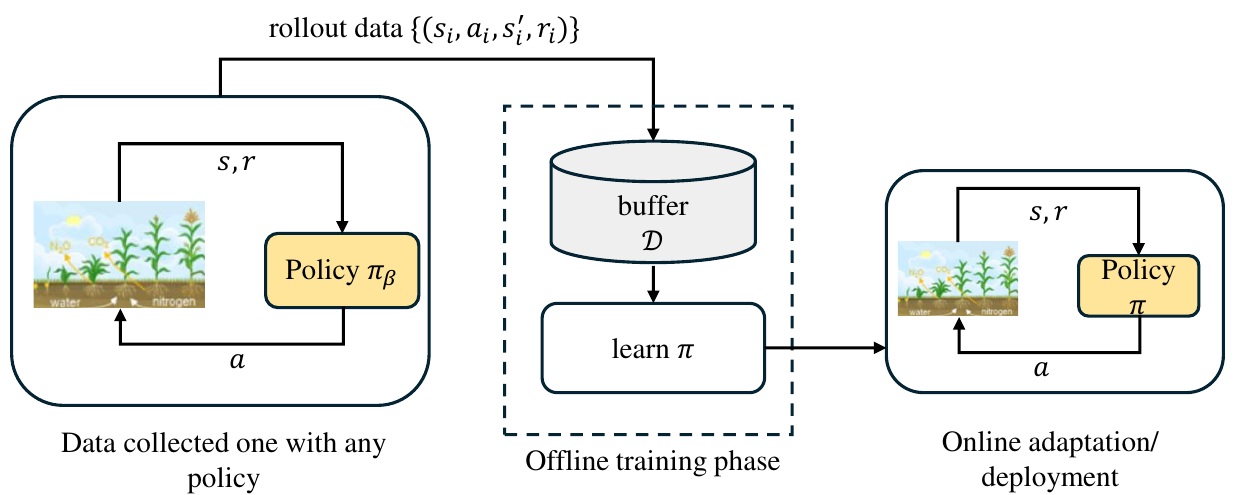}
  \vspace{1pt}
  \caption{Framework of Offline RL for crop management.
  }
  \label{fig:offlineRL}
\end{figure*}

As shown in Figure~\ref{fig:offlineRL}, the offline RL framework for crop management involves collecting data using any existing policy $\pi_{\beta}$, such as traditional crop management strategies and RL approaches, storing it in a buffer $\mathcal{D}$ during the offline training phase.
We use $\pi_{\beta}$ to denote the distribution over states and actions within $\mathcal{D}$, assuming that state-action pairs $(s, a) \in \mathcal{D}$ are sampled according to $s \sim d^{\pi_{\beta}}(s)$, with actions sampled based on the behavior policy, $a \sim \pi_{\beta}(a|s)$. Using the collected data, the optimal policy $\pi$ is learned from this static dataset. Once the policy is learned, it is deployed for online adaptation or implementation in the field. 

To demonstrate this framework, we use the batch-constrained deep Q-learning (BCQ) algorithm \citep{fujimoto2019off} as an example, which is a widely used offline RL approach. BCQ is designed to address the limitations of traditional Q-learning algorithms when applied to static datasets. The BCQ algorithm constrains the action space to be similar to the actions seen in the dataset $\mathcal{D}$, thus ensuring more reliable policy learning from offline data.

% add state representation, reward, action

In BCQ, the objective is to learn a Q-function $Q(s, a)$ and a policy $\pi(a|s)$ that maximizes the expected cumulative reward. The static dataset $\mathcal{D}$ consists of transitions $(s_t, a_t, r_t, s_{t+1})$ collected from existing policies. In the offline RL setting for crop management, the Q-function captures the long-term return of management decisions, such as when and how much to irrigate, fertilize, or control pests. Minimizing the Bellman error ensures that these decisions align closely with maximizing the reward, which could represent yield, profit, or resource efficiency. Therefore, the Bellman equation for updating the Q-function is:
\begin{equation}
Q(s_t, a_t) = r_t + \gamma \mathbb{E}_{a' \sim \pi(\cdot|s_{t+1})} \left[ Q(s_{t+1}, a') \right],
\end{equation}
where $\gamma$ is the discount factor. In practice, this is done by minimizing the Bellman error:
\begin{equation}
\begin{split}
\mathcal{L}(\theta) = & \, \mathbb{E}_{(s_t, a_t, r_t, s_{t+1}) \sim \mathcal{D}} \Bigg[ \Big( Q_\theta(s_t, a_t) - \\
& \left( r_t + \gamma \max_{a' \sim \pi(\cdot|s_{t+1})} Q_{\theta'}(s_{t+1}, a') \right) \Big)^2 \Bigg]
\end{split}
\end{equation}
where $\theta$ and $\theta'$ are the parameters of the Q-network and target Q-network, respectively.

To ensure that the policy $\pi(a|s)$ remains consistent with the behavior policy $\pi_{\beta}$, BCQ incorporates a generative model, often a variational autoencoder (VAE) \citep{kingma2019introduction}, which constrains the action space to only those actions that are likely under the behavior policy. This prevents the learned policy from making decisions that deviate significantly from those in the original dataset $\mathcal{D}$, thereby maintaining policy reliability. The VAE's role is crucial for crop management tasks where performing actions not previously observed could result in poor outcomes, such as over-irrigation, improper pest control timing, or excessive use of fertilizers. The VAE loss function is:
\begin{equation}
\begin{split}
\mathcal{L}_{\text{VAE}}(\phi) = & \, \mathbb{E}_{(s, a) \sim \mathcal{D}} \Bigg[ \|a - f_\phi(s, \epsilon)\|^2 \\
& + D_{\text{KL}} \left( q_\phi(z|s, a) \| p(z) \right) \Bigg]
\end{split}
\end{equation}
where $\phi$ are the parameters of the VAE, $f_\phi$ is the decoder, $q_\phi$ is the encoder, and $D_{\text{KL}}$ is the Kullback-Leibler divergence. The policy is then updated by maximizing the Q-value of the actions generated by the VAE:
\begin{equation}
\pi(a|s) = \arg\max_{a \sim \pi_{\beta}(a|s)} Q(s, a).
\end{equation}

By following this approach, BCQ ensures that the learned policy $\pi$ performs well when deployed in the field for crop management while adhering to the constraints imposed by the offline data.
Although BCQ is a powerful example, the offline RL framework for crop management is not limited to any single algorithm. Other algorithms, such as Conservative Q-Learning (CQL) \citep{kumar2020conservative} and implicit Q-learning (IQL) \citep{kostrikov2021offline}, could also be adapted for crop management. Each of these methods provides different strategies for learning robust policies from offline data, offering opportunities to address specific agricultural challenges, such as optimizing nutrient application schedules, minimizing disease spread, or efficiently managing labor and resources.

\subsection{LLMs for Crop Management}
\begin{figure}[!ht]
  \centering
  \includegraphics[width=0.46\textwidth]{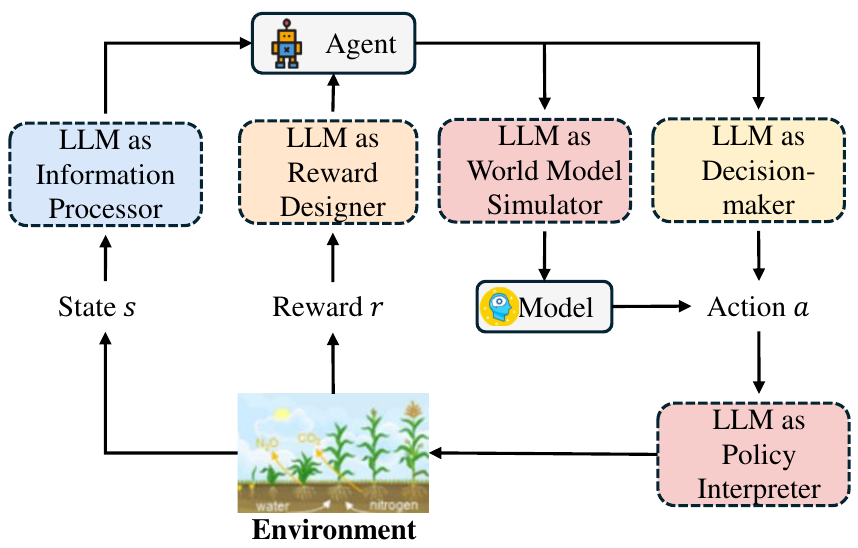}
  \vspace{1pt}
  \caption{Framework of LLM for crop management. Modified from \citep{cao2024survey}.
  }
  \label{fig:LLM}
\end{figure}

RL approaches to crop management often face challenges in designing appropriate reward functions and ensuring generalization to diverse real-world conditions. LLMs provide a promising complementary approach to address these challenges \citep{wu2024new}. As illustrated in Figure~\ref{fig:LLM}, LLMs can play multiple roles in the crop management process.

LLMs can act as powerful \textit{information processors} by extracting and interpreting state information $s$ from the crop management environment. In practical agricultural settings, observations often include complex and multimodal data, such as textual descriptions of weather forecasts, soil quality reports, or visual imagery from field sensors and drones \citep{bender2020high}. When faced with this variety of data, it becomes challenging for RL agents to simultaneously comprehend the intricacies of the information and optimize their control policy effectively. LLMs address this challenge by processing and transforming raw text and visual inputs into structured representations, which makes it easier for the agent to focus on decision-making. For example, an LLM can translate detailed weather reports or soil condition narratives into feature vectors that are more easily utilized by the agent. Similarly, visual data from field images can be analyzed by LLMs to highlight key features, such as identifying pest infestations or plant health conditions, thereby speeding up policy learning and enhancing the agent’s ability to adapt to complex, real-world agricultural tasks. Notably, \cite{wu2024new} applied LLMs as information extractors, demonstrating their potential in this domain.

Beyond processing state information, LLMs can also function as \textit{reward designers}, shaping the reward signal $r$ to better capture the multifaceted objectives of crop management. Agriculture often involves complex trade-offs, such as balancing crop yield with resource efficiency, reducing environmental impact while ensuring economic profitability, or optimizing pesticide use to minimize pest damage without compromising sustainability \citep{gautron2022reinforcement}. Designing a reward function that accurately reflects these competing priorities is crucial for effective RL-based crop management. LLMs can help in this regard by interpreting domain-specific guidelines, agronomic best practices, and expert feedback to design more nuanced and context-sensitive reward functions \citep{cao2024survey}. For example, an LLM could incorporate natural language descriptions of desired outcomes, such as ``maximize water efficiency while maintaining crop health'', and translate them into quantitative reward signals that guide policy learning. By doing so, LLMs facilitate the development of RL policies that are aligned with complex, real-world agricultural goals. 

Additionally, LLMs can act as \textit{world model simulators}, aiding policy learning by generating accurate predictions about the dynamics of the environment. In the context of crop management, modeling the environment is challenging due to the uncertainty and variability inherent in agricultural processes, such as changes in weather patterns, soil fertility, and pest behavior \citep{gautron2022reinforcement}. Traditional simulation models often require extensive domain knowledge and computational resources to build, but LLMs can simplify this process by learning to predict state transitions based on the interactions observed in historical data. For example, an LLM can be used to simulate how different irrigation strategies affect soil moisture levels over time or how varying pest control measures influence crop health. By generating these predictions, LLMs provide a simulated environment where RL agents can test and refine their policies without the risks and costs associated with real-world experimentation \citep{cao2024survey}. This approach allows for more efficient exploration of policy options and faster adaptation to changes in the crop management environment.

Furthermore, LLMs can serve as \textit{decision-makers}, directly influencing the selection of actions $a$ based on the interpreted state and policy. Traditional RL frameworks rely on learning a policy function that maps observed states to actions, often through a trial-and-error process that can be inefficient, particularly in agricultural scenarios with high uncertainty and variable conditions \citep{levine2020offline}. LLMs can streamline this process by leveraging their extensive pre-trained knowledge to suggest informed actions that align with both the current state of the environment and predefined crop management goals \citep{liu2024dellma}. For instance, upon receiving a state description indicating a weather forecast of drought and low soil moisture, an LLM decision-maker could prioritize irrigation actions over other potential options. Similarly, when the LLM identifies early signs of pest infestation from state data, it can recommend targeted pest control measures that minimize damage while aligning with sustainability goals.

Lastly, LLMs may also play a crucial role as \textit{policy interpreters}, translating actions generated by the RL agent into human-understandable recommendations that align with agronomic expertise. While RL policies are often represented as mathematical mappings from states to actions, these representations are not always intuitive or easily actionable for practitioners in agriculture \citep{bu2019smart}. LLMs can bridge this gap by converting complex policy outputs into clear, actionable guidance. For instance, instead of outputting a cryptic policy action such as ``apply nitrogen fertilizer at a rate of X kg/ha'', an LLM acting as a policy interpreter might translate this into a more comprehensive recommendation: ``Apply 50 kg/ha of nitrogen fertilizer in the early morning to optimize uptake while minimizing runoff, given the upcoming rainfall and current growth stage''. Such translations not only enhance the usability of RL policies but also ensure that recommendations are framed in a way that aligns with established agricultural best practices, ultimately supporting farmers in making informed, timely, and effective decisions.

RL faces challenges in crop management, including sample inefficiency, complex reward design, and generalization to diverse conditions \citep{gautron2022reinforcement}. Offline RL offers a promising solution by learning from existing datasets, but its effectiveness can be limited by the complexity of agricultural environments. The integration of LLMs presents significant opportunities to enhance offline RL: as information processors, they extract structured data from complex inputs; as reward designers, they align RL objectives with nuanced agricultural goals; and as world model simulators, they provide accurate predictions for policy learning. Additionally, LLMs can act as decision-makers to propose context-specific actions and as policy interpreters to translate policies into practical recommendations. Together, offline RL and LLMs can offer a robust and adaptable approach to developing effective, data-driven crop management strategies \citep{shi2023unleashing}.

\section{Summary}
\label{sec6}
Efficient and sustainable crop production process management is crucial for meeting the growing global demand for food while minimizing environmental impacts. RL and LLMs present new opportunities to enhance crop management practices, offering advanced tools for optimizing decision-making processes. However, existing approaches still face certain limitations that must be addressed for real-world application. In this paper, we investigated the recent advancements in RL and LLMs, their application within crop management DSSs, and the challenges associated with their deployment. Additionally, we explored potential solutions, such as the integration of offline RL and the use of LLMs, to address these challenges and enhance the effectiveness of crop management strategies. This study emphasizes the transformative potential of RL and LLMs in agricultural systems and the need for continued research and innovation to optimize and control crop production process management.

\section*{Authorship Contribution}
\textbf{Dong Chen}: Conceptualization, Investigation, Writing – original draft; \textbf{Yanbo Huang}: Conceptualization, Investigation, Writing – review.

% \section*{Acknowledgement}
% This work was supported in part by XX

\typeout{}
\bibliography{ref}
\end{document}